\documentstyle[twocolumn,prb,aps,graphicx]{revtex}

\ifx\pdfoutput\undefined
\def\mygraph#1#2{\includegraphics[#1]{fig#2.eps}} 
\else
\def\mygraph#1#2{\includegraphics[#1]{fig#2.pdf}} 
\fi
\def\myprefig#1{}
\def\myfigcap#1#2{}
\def\myfig#1#2{\ifpreprintsty
\else
\begin{figure}[t]
\noindent\mygraph{width=86mm}{#1}
\caption{#2}
\end{figure}\noindent
\fi}
\def\figAcaption{Complex dielectric function $\varepsilon_1+{\rm i}\varepsilon_2$ of Yb-123
at room temperature. The inset depicts the Raman cross-section
for the x'x' (top panel) and the x'y' (bottom panel) polarization
symmetries. Lines mark the average values.\label{fig:bgint}}

\def\figBcaption{Raman spectra of Yb-123 in x'x' and x'y' polarization geometries taken with
excitation energies $\hbar\omega_i=2.71$ eV (dashed lines) and
1.96 eV (solid) at 15~K cryostat temperature.
The lower panels show the background spectra obtained by subtraction of the
phonons.\label{fig:specs}}

\def\figCcaption{Temperature dependence of the integrated O(4) Raman intensity
in Yb-123 in x'x' (closed symbols) and x'y' (open symbols) polarizations
for excitation energies 2.71 eV (triangles), 2.41 eV (diamonds), 1.96 eV (circles), and 1.68 eV (squares).
Solid lines serve as guides to the eye, dashed lines indicate $T_c$.
The x'y' data are offset as indicated.\label{fig:o4temp}}

\def\figDcaption{O(4) intensity of Yb-123 at 15 K in x'x' (closed symbols) and x'y' (open
symbols) polarization geometry. Lines are guides to the eye.\label{fig:o4resenh}
The inset shows the ratio of the x'x' and x'y' intensity at 15 K
for various $R$-123 single crystals that correspond to different doping levels.}

\def\figEcaption{Temperature dependence of the O(2)-O(3) intensity of Yb-123 for excitation
energies $\hbar\omega_i=$~2.71 eV (triangles), 2.41 eV (diamonds), 1.96 eV (circles), and 1.68 eV (squares).
The dotted line indicates $T_c$.\label{fig:o2o3}}

\def\figFcaption{Resonance of the intensity of the Cu(2) mode in Yb-123 at 15 K in x'x' (closed symbols) and
x'y' (open symbols) geometry. Dotted Gaussian profiles guide the eye.\label{fig:cu}}

\begin{document}
\twocolumn[\hsize\textwidth\columnwidth\hsize\csname@twocolumnfalse\endcsname

\preprint{Submitted to PR{\bf B}}

\draft

\title{Superconductivity-induced changes of the phonon resonances in $R\bf Ba_2Cu_3O_7$ ($R$ = rare earth)}

\author{S. Ostertun, J. Kiltz, A. Bock\cite{AB}, and U. Merkt}

\address{Institut f\"{u}r Angewandte Physik und Zentrum f\"{u}r
         Mikrostrukturforschung, \\
         Universit\"{a}t Hamburg, Jungiusstra{\ss}e 11, D-20355 Hamburg,
         Germany}

\author{T. Wolf}

\address{Institut f\"{u}r Festk\"{o}rperphysik,
         Forschungszentrum Karlsruhe, D-76021 Karlsruhe, Germany}

\date{\today}

\maketitle

\begin{abstract}
We observe a characteristic energy $\hbar\omega_{c}\approx 2.1$ eV which
separates regions of different behavior of the phonon intensities
in the Raman spectra of the $R\rm Ba_2Cu_3O_7$ system.
A superconductivity-induced drop of phonon intensities is found
for the oxygen modes O(4) and O(2)-O(3) only for excitation
energies below $\hbar\omega_c$.
This intensity drop indicates an order parameter which affects 
energies in the vicinity of $\hbar\omega_c$.
\end{abstract}

\pacs{PACS numbers: 74.25.Gz, 74.25.Jb, 74.25.Kc, 74.72.Bk, 78.30.Er}

]  

\narrowtext

\section{introduction}
Several superconductivity-induced changes of parameters describing
the phonons in the cuprate superconductors can be determined by
Raman spectroscopy.
Frequency and linewidth anomalies like hardening and broadening
have been the subject of several experimental\cite{Altendorf:93,Hadjiev:98,Bock:99PRB} and
theoretical\cite{Zeyher:90,Nicol:93,Devereaux:94,Normand:96} studies.
While the superconductivity-induced phonon self-energy effects allow
conclusions regarding the order parameter only for
energies similar to the phonon energies (up to 100 meV),
high-energy ($>$1 eV) questions cannot be
addressed this way.
Resonant Raman scattering is an experimental technique which
allows one to examine the physics at high energy through the resonances of
phonons. In contrast to reflectivity or transmission spectroscopy the affected
phonons provide direct information on incorporated electronic bands as the
assignments of the phonon modes in the spectra to the vibrating atoms is well
known through group theoretical calculations and isotope
substitution experiments.
Thus, the dependence of the phonon intensities on
temperature and excitation energy can provide information about the
scattering mechanism and especially about the superconducting state.

In contrast to the usually observed increase of the
intensity\cite{Sherman:95,Misochko:99} below the critical
temperature $T_c$ we observe a drop of intensity in several modes
when exciting with photons of energy $\hbar\omega_i<2.1$~eV.
In the same region of excitation energy the apical oxygen mode at
500 cm$^{-1}$ shows a violation of the symmetry selection rule
which cannot be explained in terms of an orthorhombic distortion.
Summarizing our data we conclude that the origin of the intensity
anomaly is related to a modification of the Cu-O charge transfer
mechanism in the superconducting state in
comparison to the normal state.

\section{experimental details}
Subject of this paper are Yb-123, Er-123, Sm-123 and Nd-123 single
crystals, with $T_c=$ 76~K, 81~K, 94~K,
and 90~K, respectively.
All measured single crystals were grown with a self-flux method and
annealed with oxygen under high pressure.\cite{Wolf:89}
Due to the high oxygen content and the different radii of the rare earth atoms
Yb-123 and Er-123 are overdoped whereas Sm-123 and Nd-123
are nearly optimally doped.\cite{Xu:92}
The laser beam is focused onto the sample along the c-direction.
The orthorhombic symmetry of the $R$-123 system can be treated as
tetragonal. Accordingly, the Raman active phonons are of
$\rm A_{1g}$ and $\rm B_{1g}$ symmetry,
which are allowed for $\rm z(xx)\bar{z}$/$\rm z(x'x')\bar{z}$ and
$\rm z(xx)\bar{z}$/$\rm z(x'y')\bar{z}$ polarizations in
the Porto notation, respectively.
All polarizations are specified with respect to the axes along the
Cu-O bonds of the $\rm CuO_2$ planes, primed polarizations are rotated
by $\rm 45^o$. For simplicity z and $\rm \bar{z}$
are omitted in the following.
The measurements were performed using several lines of
Ar$^+$, He-Ne and Ti:Saphir lasers in quasi-backscattering geometry.
The details of the setup are described elsewhere.\cite{Ruebhausen:97b}
In order to achieve a high accuracy of intensity measurements the
laser power was monitored during the measurements.
All spectra were corrected for the response of the detector and the
optical system. Also, they are normalized to the incident photon rate.

For a comparison of intensities obtained with different excitation
energies the cross-section is calculated from the efficiencies
using ellipsometric data of Yb-123.
The complex dielectric function of Bi-2212 shows only a slight
variation with temperature\cite{Ruebhausen:01} at our excitation
energies between 1.68 eV and 2.71 eV and the resulting reflectivity
exhibits maximal variations
below 1\%. As the reflectivity of the $R$-123 system behaves
similar,\cite{Holcomb:96}
Raman spectra of all temperatures can be
evaluated with the ellipsometric data obtained at room-temperature.
In Fig.\ \ref{fig:bgint} the complex dielectric function of Yb-123
at room-temperature is given. From the real part $\varepsilon_1$
of the dielectric function
we can estimate a screened plasma frequency $\omega_p$ of 1.44 eV.

The low-temperature background intensity of the
cross-section at $\omega\approx 700~\rm
cm^{-1}$ is plotted versus the excitation energy for the x'x' and x'y'
polarization geometry in the inset of Fig.\ \ref{fig:bgint} in the top and
bottom panel, respectively.
Within experimental error no significant
resonance of the cross-section can be determined. The scattering
of the data points results from uncertainties in the adjustment of the setup.
Thus, the measured phonon intensities will scatter in a similar
way.
To obtain the integrated phonon intensities we fitted the spectra
to the model presented in a previous paper.\cite{Bock:99PRB}
An extended Fano profile is described by
\begin{eqnarray}
I(\omega)&=&\frac{C}{\gamma(\omega)\left[1+\epsilon^2(\omega)\right]}\times\nonumber\\
&&\left\{\frac{R_*^2(\omega)}{C^2}-2\epsilon(\omega)\frac{R_*(\omega)\varrho_*(\omega)}{C^2}-\frac{\varrho_*^2(\omega)}{C^2} \right\}
\end{eqnarray}
with the substitutions
$\varrho_*(\omega)=Cg_{\sigma}^2\varrho_{\sigma}^e(\omega)$,
$R_*(\omega)=Cg_{\sigma}^2R_{\sigma}^e(\omega)+R_0$,
$\epsilon(\omega)=\left[\omega^2-\omega_{\nu}^2(\omega)\right]/2\omega_p\gamma(\omega)$,
$\gamma(\omega)=\Gamma+\varrho_*(\omega)/C$, and
$\omega_{\nu}^2(\omega)=\omega_p^2-2\omega_pR_*(\omega)/C$.
Here, $\varrho_{\sigma}^e(\omega)$ and $R_{\sigma}^e(\omega)$ are the imaginary
and real part of the electronic response function, respectively,
and $g_{\sigma}$ is the lowest-order expansion coefficient of the
electron-phonon vertex. The bare frequency and linewidth of the
phonon are $\omega_p$ and $\Gamma$, respectively. The constant $C$ is a
scaling parameter due to the use of arbitrary units in the spectra.

We use this extended Fano profile for the asymmetric phonon modes
and Lorentzian profiles for the remaining
phonons.
As described in Ref.\ \onlinecite{Bock:99PRB},
the measured electronic response $\varrho_*(\omega)$ is described via
a phenomenological formula which contains a ${\rm tanh}(\omega/\omega_T)$ term for
the incoherent background and
two coupled Lorentzian profiles
for the redistribution of the spectra below $T_c$ --- one for
the pair breaking peak and a negative one for the suppression of
spectral weight at low Raman shifts. This formula allows a
simultaneous description of $R_*(\omega)$.

\section{results}
In the top panels of Fig.\ \ref{fig:specs} we present
spectra of Yb-123 obtained at 15~K nominal
temperature in x'x'
($\rm A_{1g}+B_{2g}$) and in x'y' ($\rm B_{1g}$) geometry
with excitation energies of 2.71 eV and 1.96 eV.
The original spectra have been scaled to equal background intensity for
clarity. Exciting with 2.71 eV all phonons show up mainly in the spectra of
the expected polarization geometry.
For the x'x' geometry we have the Ba mode at 120 cm$^{-1}$,
Cu(2) at 150 cm$^{-1}$, O(2)+O(3) at 440 cm$^{-1}$, and O(4) at 500
cm$^{-1}$. For the x'y' geometry the O(2)-O(3) mode appears at 340 cm$^{-1}$.
In contrast, exciting with 1.96 eV the O(4) mode has nearly vanished in
x'x' geometry but only slightly changed in x'y' symmetry.
Nearly the same behavior is observed for the resonance of the intensity of the O(4) mode
in Er-123, Sm-123, and Nd-123.
Orthorhombic distortion, which is usually mentioned as the cause for phonon intensity in
forbidden symmetries, cannot explain this behavior.
Though resonance effects can disturb symmetry selection rules in
principle, the symmetry violation should appear under resonance
conditions but not out of resonance.

Subtracting the phononic contribution we get the electronic
background of Yb-123, which is plotted in the lower panels of
Fig.\ \ref{fig:specs}.
The x'x' spectra are nearly identical for both excitation energies and even
the x'y' spectra are quite similar.
The $\rm B_{1g}$ pair breaking peak lies at a relatively
low energy of $\rm 320\rm\,cm^{-1}$ due to the high doping
level\cite{Bock:99AdP} of $p\approx 0.2$, which is determined
from $T_c/T_{c,\rm max}$ (see Ref. \onlinecite{Tallon:95}).
There is no enhancement of the gap excitation from 1.96 eV to 2.71
eV. This is consistent with the behavior observed for Bi-2212
which shows non-resonant gap excitations for doping levels
up to approximately 0.2 and a gap resonance
only for higher doping levels
as reported in Ref. \onlinecite{Ruebhausen:99}.

Figure \ref{fig:o4temp} depicts the results of a detailed study
of the integrated O(4) mode intensity.
For temperatures above $T_c$ the x'x' intensity
increases when the temperature is lowered for excitation
energies of 1.96 eV, 2.41 eV, and 2.71 eV.
This increase is hardly influenced by the phase transition below
$T_c$ for 2.41 eV and 2.71 eV in contrast to 1.96 eV where the
intensity suddenly drops.
The comparatively high values for intensities of the 2.71 eV data
in the range of 100 K to 200 K are due to the scattering
mentioned in the introduction. Here it results
from a slight shift of the laser spot with respect to the entrance
slit of the spectrometer.
On the other hand the x'y' intensity
shows the same behavior for all measured excitation
energies: a slight increase with decreasing
temperature not or only slightly affected by $T_c$.
For temperatures above
$T_c$ the relation for the x'x' intensities
$I_{2.71}$:$I_{2.41}$:$I_{1.96}$ is approximately 5.0:3.1:1.0 as
indicated by the solid lines
in the left panel of Fig.\ \ref{fig:o4temp}.
The superconductivity-induced drop of the x'x' intensity at
low excitation energies
leads to a superconductivity-induced enhancement
of the resonance profile of the x'x' O(4) mode intensity.

Figure \ref{fig:o4resenh} yields the excitation energy
$\hbar\omega_c=2.1\pm 0.1\rm~cm^{-1}$ where the
O(4) mode exhibits equal intensity in x'x' and x'y' geometries.
This is also the energy of the crossover
between the superconductivity-induced increase and decrease
of the x'x' intensity shown in Fig.\ \ref{fig:o4temp}, i.\ e.\ 2.0-2.4 eV.
While the measurements have been carried out in this detail for Yb-123
only, spectra of Er-123, Sm-123, and Nd-123 at low temperature
exhibit a similar behavior as shown in the inset of
Fig.\ \ref{fig:o4resenh}. There the ratio of the x'x' and x'y' O(4) mode intensity
is plotted against the excitation energy. It exhibits no significant
doping dependence.
The O(2)-O(3) mode displays a similar behavior in the x'y' as the O(4)
mode in the x'x' symmetry. In Fig.\ \ref{fig:o2o3} the temperature dependence
of its intensity is plotted for excitation
energies $\hbar\omega_i=$ 1.68 eV, 1.96 eV, 2.41 eV, and 2.71 eV.
Above $T_c$ the intensity is only slightly
affected by the temperature, below $T_c$ the
intensity drops only for excitation energies below $\hbar\omega_c$.

The intensity of the Cu(2) mode shows a strong increase with
decreasing temperature above $T_c$ but also a slight drop below $T_c$.
The intensity at low temperatures is approximately 70\%--80\% of the highest
value at $T_c$.
While this ratio is independent of excitation energy and polarization geometry
the resonance energy itself depends on the polarization geometry
as shown in Fig.\ \ref{fig:cu}.
Gaussian profiles guide the eye and help to determine the
resonance energy of the Cu(2) mode intensity, which is
about 2.1 eV for x'x' and well below 2.0~eV for x'y' symmetry.
The Ba mode which is not shown here
exhibits a slight
intensity gain below $T_c$, which seems to be
unaffected by $\hbar\omega_c$.
The x'x' intensity is resonant at 2.6 eV or even higher energy,
the x'y' intensity follows this profile with about 30\% of the intensity
in x'x' geometry.
\section{discussion and conclusions}
Summarizing our data we find a characteristic energy
$\hbar\omega_c\approx2.1\rm~eV$ which separates the regions of
red and blue excitations. It is very similar to the crossing point
at 2.2 eV observed in the imaginary part of the dielectric
function\cite{Ruebhausen:01} of Bi-2212.
For red excitation the $R$-123 system exhibits
a superconductivity-induced drop of the intensity of the O(4)
mode in x'x' and the O(2)-O(3) mode in x'y' symmetry, for blue excitation the
drop is absent. Additionally, for red excitations the symmetry selection rule
of the O(4) mode is violated as the x'y' intensity exceeds the x'x'
intensity. The maximum of the intensity of the
Cu(2) mode in x'x' symmetry coincides
with $\hbar\omega_c$.
All these effects are characterized
by $\hbar\omega_c$ indicating a common microscopic origin.
Low temperature measurements of overdoped Er-123 as well as of
underdoped Sm-123 and Nd-123 show a
similar behavior as Yb-123, especially the symmetry
violation of the O(4) mode for excitation energies below
$\hbar\omega_c$.
Thus $\hbar\omega_c$ could be a general property of the $R$-123
system without a significant doping dependence.

Friedl {\em et al.}\cite{Friedl:91} observed a
superconductivity-induced gain of the xx phonon intensities of Y-123 films
on $\rm SrTiO_3$ substrates, which are independent
of excitation energy in the range of 1.92~eV to 2.60~eV,
a drop of intensity is observed only for the Cu(2) mode.
The differences between
their data and the data presented here
may in principle result from the different doping levels
of Y-123 and Yb-123. But as we have noted above, the effects are
mainly
independent of doping,
our explanation for the discrepancies between Friedl's
and our data is the following:
As the lattice constant of
Y-123 (3.82/3.88 \AA) is smaller than that of $\rm SrTiO_3$ (3.91
\AA) and as the thermal expansion coefficient
of Y-123 (11.7 \AA/K) exceeds that of $\rm SrTiO_3$
(9.4 \AA/K), the
lattice mismatch increases with decreasing temperature. The
additional stress may explain the high intensity
increase I(100K):I(250K)$\approx$2.5 of the Y-123 films in
comparison to I(100K):I(300K)$\approx$1.3 of the Yb-123 single crystal.
Thus, the drop below $T_c$ is masked in this strong increase, especially
as Friedl {\em et al.}\ recorded xx+yy data which include x'y' where no intensity
drop appears.
In underdoped $R$-123 the O(2)-O(3) mode is known to show a strong
intensity gain.\cite{Bock:99PRB} Thus, the intensity drop of the O(2)-O(3) mode can
only be observed in overdoped samples as
the drop is not masked by an intensity gain.

Sherman {\em et al.}\cite{Sherman:95} explained the increase of the
total $\rm B_{1g}$ mode intensity below $T_c$ of Y-123
in terms of an extension of the number of intermediate
electronic states near the Fermi surface
that participate in the Raman process.
Accordingly, the doping dependence of the gain of the O(2)-O(3) mode intensity can
be explained in this picture by the vicinity to the Fermi-level.
Analogously, an intensity drop means a reduction of the
number of intermediate states which is not consistent with this
model. Possibly the intensity gain and the intensity drop
have different origins, which are doping dependent and
independent, respectively.

Due to the different resonance profiles in x'x' and x'y' geometry
the drop of the phonon intensities
can only be explained in the framework of a resonant theory
which must account for the bands of the initial and final states.
Calculations for the phonons of Y-123 in the superconducting state which have been performed by
Heyen {\em et al.}\cite{Heyen:90} and which are based on the local-density
approximation and the linear muffin-tin-orbital method
do not show any signature of an intensity drop around 2 eV for xx/yy and zz symmetry.
The calculated xx/yy intensity of the O(4) mode has a local intensity minimum at
2.0 eV but at 1.6 eV the predicted intensity is more than twice
the intensity at 2.0 eV excitation energy which is strong contrast to our data.

As no appropriate theory is available we try an
interpretation in a simple picture: The intensity drop is observed
for the Cu(2) mode and, for excitation
energies below $\hbar\omega_c$, for the plane-related O(2)-O(3)
mode and for the x'x' O(4) mode.
This indicates that plane bands are responsible for the observed
resonance effect. In case of the O(4) mode two processes
seem to contribute to the phonon intensity, as the strongly resonant x'x'
intensity exhibits the drop below $\hbar\omega_c$ and the
x'y' intensity shows no resonance and no superconductivity-induced
effects.
As the characteristic energy $\hbar\omega_c$ which separates
the regions with and without the superconductivity-induced
intensity drop is the same for the O(4) mode as for the
O(2)-O(3) mode, we conclude that the relevant initial or final
state bands are the same. Thus, we
attribute the x'x' intensity to a process involving that plane
bands.
On the other hand the x'y' intensity is attributed to the chain
bands. Another hint for this interpretation is that the
pairing mechanism is believed to be located in the
$\rm CuO_2$ planes. Thus, superconductivity-induced effects
would be expected for processes which involve plane bands.

Within a simple band structure picture the opening of the gap
cannot explain the threshold of $\hbar\omega_c$ directly:
If the opening of the gap would enhance the energy distance
between the possible initial and final states for an allowed
transition to a value which exceeds the excitation energy, this
process would also be suppressed for lower excitation energies
in the normal state. Figure \ref{fig:o2o3} clearly shows that the
behavior of the phonon intensity for 1.96 eV and 1.68 eV excitations
is essentially the same. Thus, only a fundamental change of the
band structure could explain the different behavior for blue and
red excitation.
As the x'x' intensity of the Cu(2) mode has its
maximum right at the critical energy $\hbar\omega_c$
(see Fig.\ \ref{fig:cu}) the fundamental change
with the superconducting transition
should take place in a plane band involving the Cu(2) atom.
As mentioned above the O(4) mode intensity in x'x' and x'y'
symmetry results from different processes due to the coupling to the
Cu(2)-O(2)-O(3) plane bands and to the Cu(1)-O(1) chain bands,
respectively.

In comparison to the superconducting gap $2\Delta$ the critical
energy $\hbar\omega_c$ is quite a large energy.
Superconductivity-induced high-energy effects have been observed
previously by thermal difference reflectance (TDR)
spectroscopy\cite{Holcomb:96} and
by ellipsometry.\cite{Ruebhausen:01}
In the ellipsometry data a crossing point in $\varepsilon_2$
at 2~eV separates two spectral regions of different behavior in
their temperature dependencies.
The TDR spectra of Y-123 and other high-temperature superconductors
exhibit a deviation from
unity in the ratio $R_S/R_N$ of the superconducting to normal
state spectra at high photon energies ($\approx$2.0 eV).
Within the Eliashberg theory the deviation can only be explained
if the electron-boson coupling function contains a high-energy
component in addition to the electron-phonon interaction leading to
an order parameter which is non-zero for similar energies.
The microscopic origin of the high-energy interaction
is most likely a $d^9-d^{10}\underline{L}$ Cu-O charge-transfer
within the $\rm CuO_2$ planes.\cite{Holcomb:96}

The Cu(2) mode intensity seems to correlate with the charge-transfer
excitation at about 2.1 eV. The intensity drop of this mode below
$T_c$ is independent of excitation energy and supports the
interpretation of a modified charge-transfer-mechanism in the
superconducting state.
On the other hand the O(2)-O(3) mode in x'y' and the O(4) mode in
x'x' geometry behave completely different for excitation energies
above and below $\hbar\omega_c$. This indicates that the
underlying mechanism is not the modified charge-transfer itself
but the opening of the gap, which is non-zero for high energies
as argued by Holcomb {\em et al.}\cite{Holcomb:96}
Thus, superconductivity-induced high-energy effects
which cannot be detected in transport
experiments, leave their signature in resonant Raman experiments
through different behavior of the phonon intensity above and below $\hbar\omega_c$
as well as in other optical experiments.\cite{Ruebhausen:01,Holcomb:96}

\section{Summary}
A resonance enhancement of the integrated phonon intensity is
observed for the O(4) mode in x'x' and for the O(2)-O(3) mode in x'y' polarization geometry.
It results mainly from an intensity drop for temperatures below $T_c$
at excitation energies less then $\hbar\omega_c\approx 2.1$ eV.
This drop is also observed for the Cu(2) mode, which has its maximum
x'x' intensity right at this excitation energy.
We believe that the charge-transfer is responsible for the
intensity drop of all three phonon modes, directly for Cu(2) and
indirectly via the order parameter for O(4) and O(2)-O(3).
As the x'y' intensity of the O(4) mode shows hardly any resonance
we believe that its origin is completely different from the
one of the x'x' intensity.
\acknowledgements
The authors thank M. R\"{u}bhausen and M. V. Klein for stimulating discussions and
for performing the ellipsometry measurements
of Yb-123.
Financial support of the Deutsche Forschungsgemeinschaft
via the Graduiertenkolleg
``Physik nanostrukturierter Festk\"{o}rper''
is gratefully acknowledged.

\clearpage
\myfig{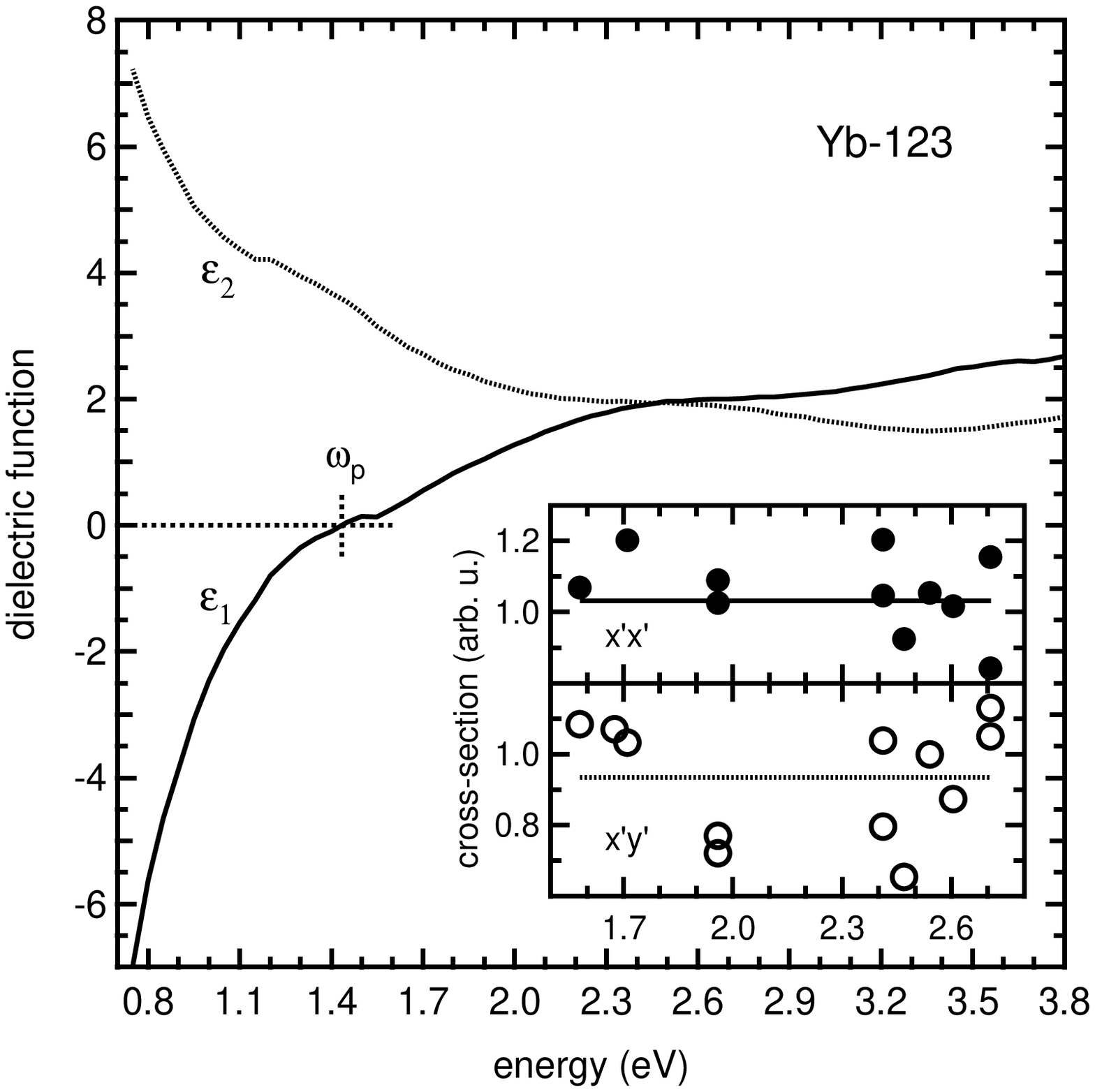}{\figAcaption}\noindent
\myfig{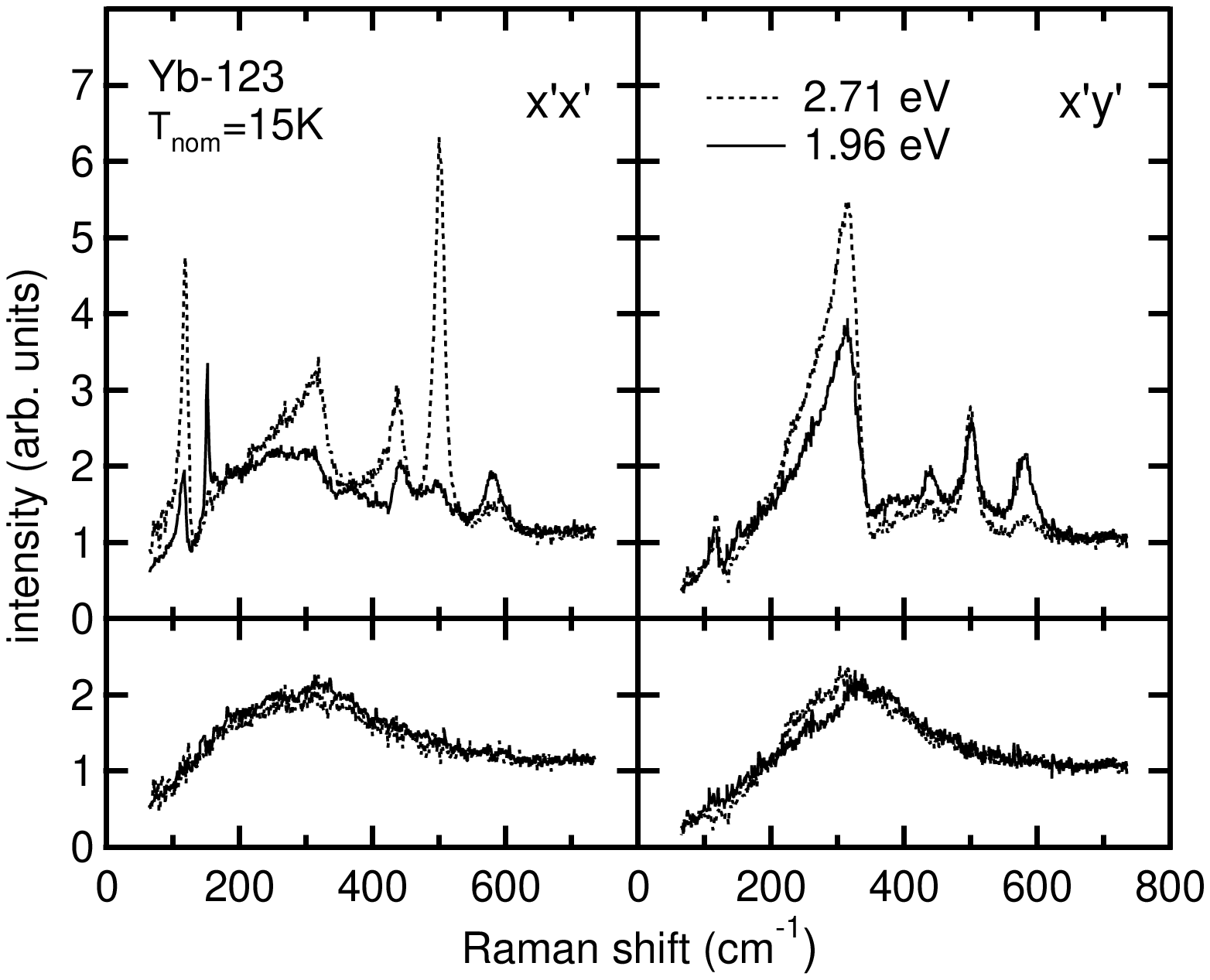}{\figBcaption}\noindent
\myfig{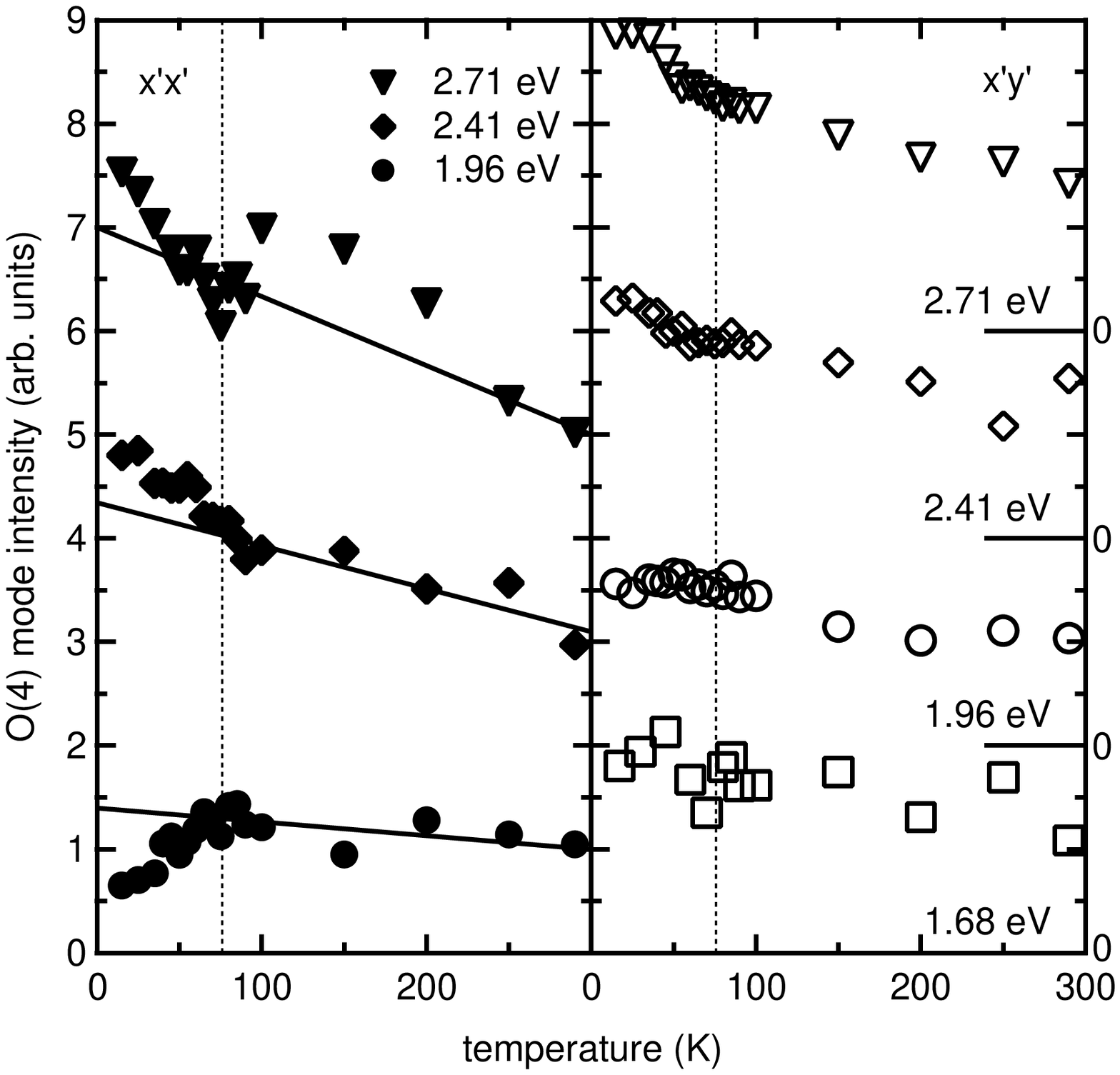}{\figCcaption}\noindent
\myfig{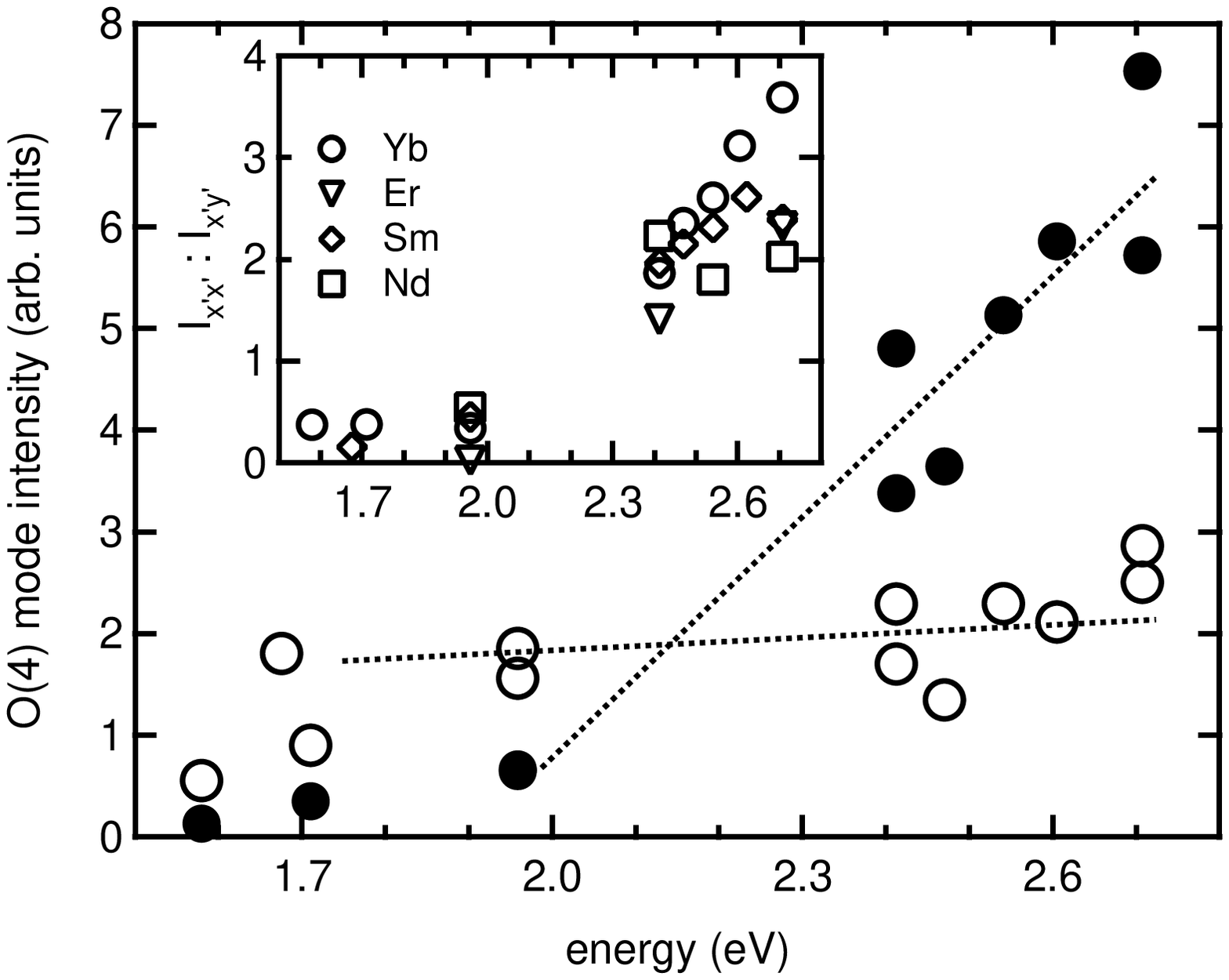}{\figDcaption}\noindent
\myfig{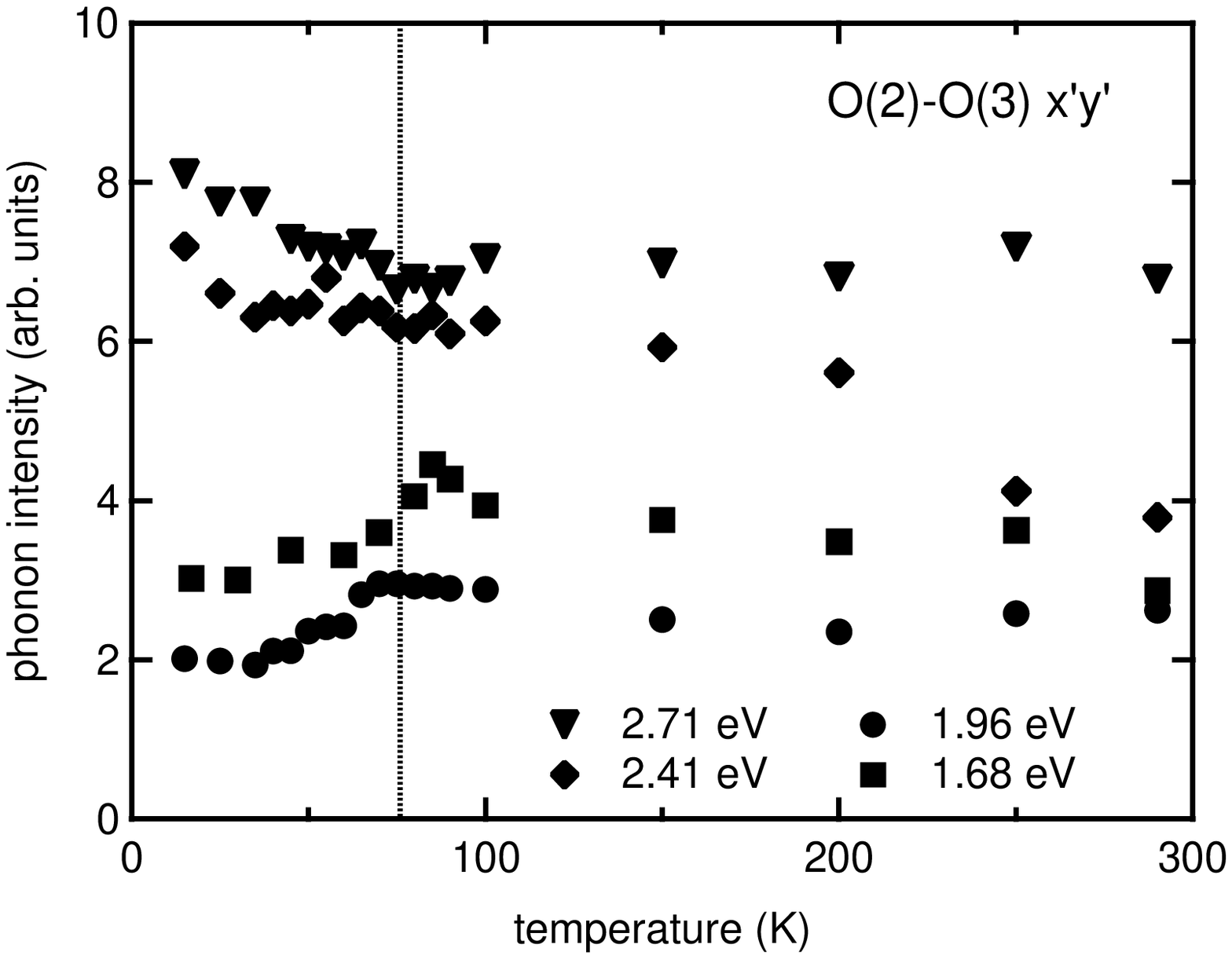}{\figEcaption}\noindent
\myfig{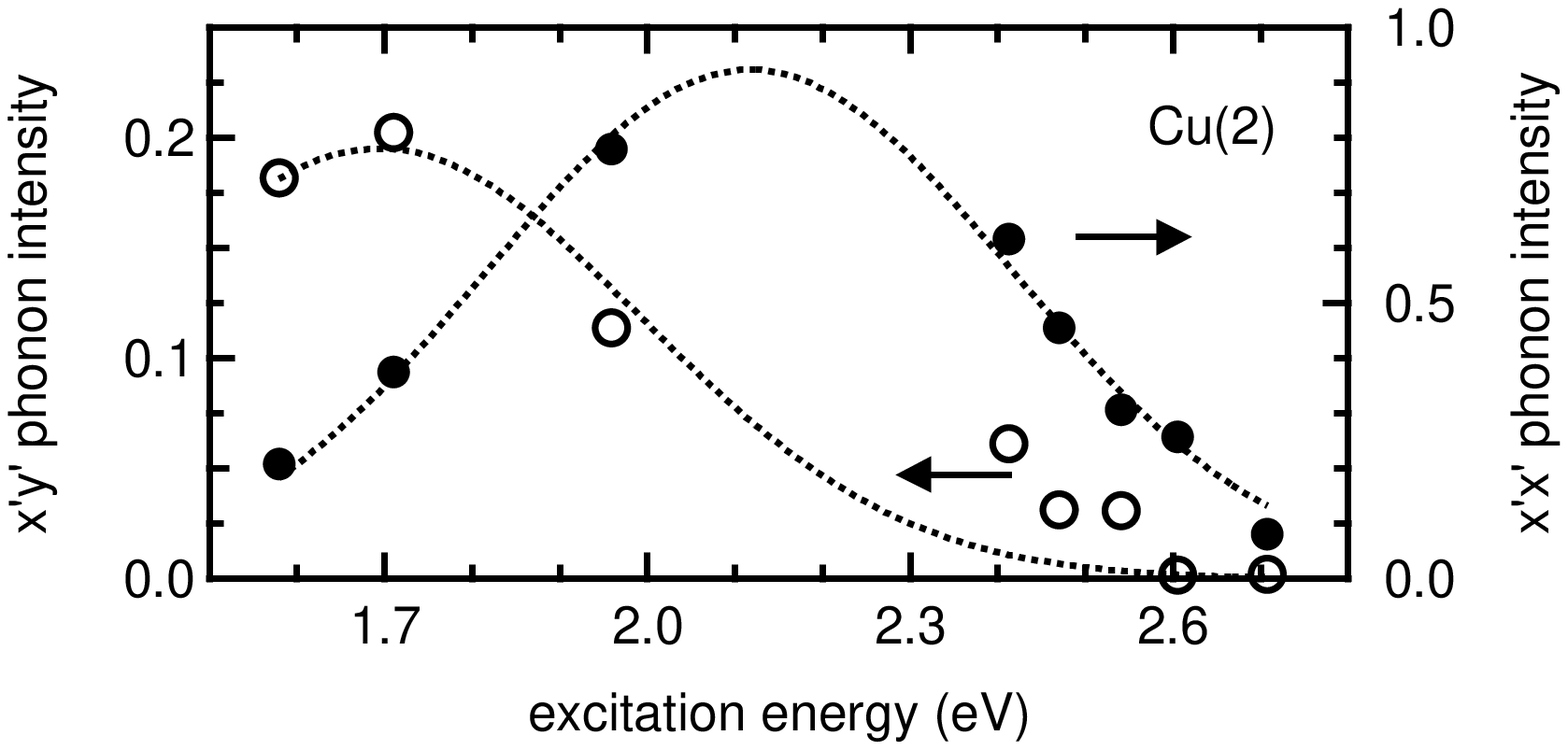}{\figFcaption}\noindent


\begin{references}
\bibitem[*]{AB}Present Address: Basler AG, An der Strusbek 60-62,
D-22926 Ahrensburg, Germany.

\bibitem{Altendorf:93}
E.~Altendorf, X.~K. Chen, J.~C. Irwin, R.~Liang, and W.~N. Hardy,
\newblock {Phys. Rev. B} {\bf 47}, 8140, 1993.

\bibitem{Hadjiev:98}
V.~G. Hadjiev, {Xingjiang Zhou}, T.~Strohm, M.~Cardona, Q.~M. Lin, and C.~W.
Chu,
\newblock {Phys. Rev. B} {\bf 58}, 1043, 1998.

\bibitem{Bock:99PRB}
A.~Bock, S.~Ostertun, R.~Das Sharma, M.~R\"ubhausen, and K.-O.
Subke,
\newblock {Phys. Rev. B} {\bf 60}, 3532, 1999.

\bibitem{Zeyher:90}
R. Zeyher and G. Zwicknagel,
\newblock {Z. Phys. B} {\bf 78}, 175, 1990.

\bibitem{Nicol:93}
E.~J. Nicol, C. Jiang, and J.~P. Carbotte,
\newblock {Phys. Rev. B} {\bf 47}, 8131, 1993.

\bibitem{Devereaux:94}
T.~P. Devereaux,
\newblock {Phys. Rev. B} {\bf 50}, 10287, 1994.

\bibitem{Normand:96}
B. Normand, H. Kohno, and H. Fukuyama,
\newblock {Phys. Rev. B} {\bf 53}, 856, 1996.

\bibitem{Sherman:95}
E. Ya. Sherman, R. Li, and R. Feile,
\newblock {Phys. Rev. B} {\bf 52}, R15757, 1995.

\bibitem{Misochko:99}
O.~V. Misochko, E.~Ya. Sherman, N.~Umesaki, K.~Sakai, and
S.~Nakashima,
\newblock {Phys. Rev. B} {\bf 59}, 11495, 1999.

\bibitem{Wolf:89}
T.~Wolf, W.~Goldacker, B.~Obst, G.~Roth, and R.~Flukiger,
\newblock {J. Crystal Growth} {\bf 96}, 1010, 1989.

\bibitem{Xu:92}
Y.~Xu and W.~Guan,
\newblock {Phys. Rev. B} {\bf 45}, 3176, 1992.

\bibitem{Ruebhausen:97b}
M.~R\"ubhausen, C.~T. Rieck, N.~Dieckmann, K.-O. Subke, A.~Bock, and
U.~Merkt,
\newblock {Phys. Rev. B} {\bf 56}, 14797, 1997.

\bibitem{Ruebhausen:01}
M. R\"{u}bhausen, A. Gozar, M.~V. Klein, P. Guptasarma, and D.~G. Hinks,
\newblock {submitted to Phys. Rev. B}

\bibitem{Holcomb:96}
M.~J. Holcomb, C.~L. Perry, J.~P. Collman, and W.~A. Little,
\newblock {Phys. Rev. B} {\bf 52}, 6734, 1996.

\bibitem{Bock:99AdP}
A.~Bock,
\newblock {Ann. Phys.} {\bf 8}, 441, 1999.

\bibitem{Tallon:95}
J.~L. Tallon, C.~Bernhard, H.~Shaked, R.~L. Hitterman, and J.~D.
Jorgensen,
\newblock {Phys. Rev. B} {\bf 51}, 12911, 1995.

\bibitem{Ruebhausen:99}
M.~R\"ubhausen, O.~A. Hammerstein, A.~Bock, U.~Merkt, C.~T. Rieck,
P.~Guptasarma, D.~G. Hinks, and M.~V. Klein,
\newblock {Phys. Rev. Lett.} {\bf 82}, 5349, 1999.

\bibitem{Friedl:91}
B. Friedl, C. Thomsen, H.-U. Habermeier, and M. Cardona,
\newblock {Solid State Commun.} {\bf 78}, 291, 1991.

\bibitem{Heyen:90}
E.~T. Heyen, S.~N. Rashkeev, I.~I. Mazin, O.~K. Andersen, R.~Liu, M.~Cardona, and
O.~Jepsen,
\newblock {Phys. Rev. Lett.} {\bf 65}, 3048, 1990.

\end{references}
\end{document}